# Advanced noise reduction techniques for ultra-low phase noise optical-to-microwave division with femtosecond fiber combs.

W. Zhang[1], Z. Xu[2], M. Lours[1], R. Boudot[2], Y. Kersalé[2], A.N. Luiten[3], Y. Le Coq[1]

and G. Santarelli[1]

[1]*LNE-SYRTE, Observatoire de Paris, CNRS, UPMC, 61, Avenue de l'Observatoire,*

*Paris, France*

[2]*FEMTO-ST Institute, CNRS and ENSMM, Besançon, France*

[3]*School of Physics, University of Western Australia, Crawley 6009, Australia*

## Abstract

We report what we believe to be the lowest phase noise optical-to-microwave frequency division using fiber-based femtosecond optical frequency combs: a residual phase noise of -120dBc/Hz at 1 Hz offset from a 11.55GHz carrier frequency. We furthermore report a detailed investigation into the fundamental noise sources which afflicts the division process itself. Two frequency combs with quasi-identical configurations are referenced to a common ultrastable cavity laser source. To identify each of the limiting effects we implement an ultra-low noise carrier-suppression measurement system, which circumvents the detection and amplification noise of more conventional techniques. This technique now allows the suppression of these unwanted sources of noise to very low levels. In the Fourier frequency range of ~200 Hz to 100 kHz, a fed-forward technique based on a voltage-controlled phase shifter delivers a further noise reduction of 10 dB. For lower Fourier frequencies, optical power stabilization is implemented in order to reduce the relative intensity noise which causes unwanted phase noise through power to phase conversion in the detector. We implement and compare two possible control schemes based on an acousto-optical modulator and comb pump current. We also present wideband measurements on the relative intensity noise of the fiber comb.

*1.Introduction.*

Low phase noise microwave signals find use in critical technological applications such as radar, telecommunications, deep space navigation systems, timing distribution and synchronization [1] as well as in the leading edge of science such as Very Long-Baseline Interferometry and development of local oscillators for accurate fountain atomic frequency standards [2-5]. The very lowest phase noise signals of –100dBc/Hz at 1Hz offset from the carrier have been generated by Cryogenic Sapphire Oscillators (CSO) based on high-quality factor whispering gallery mode sapphire resonators [4,5]. However, the CSOs are complex laboratory-based devices that are highly unlikely to be deployable to meet wider technological needs. It is thus essential to continue development of novel and less complex low phase-noise sources, which can equal or surpass the CSO performance, so as to meet these technological requirements.

One very promising alternative approach is the use of ultra-stable lasers together with the use of a frequency comb as a low-noise frequency divider to generate low phase noise microwave signals [6-11]. The low noise of frequency combs has already provided a revolution in optical metrology field where it has been used for optical frequency measurements with uncertainty close to the limitations of the caesium fountain clock [12-15] as well as frequency comparisons between different optical clocks [16-18]. In parallel, recent years have witnessed the development of compact and potentially portable, vibration-insensitive and stable optical cavities which allow fractional frequency stabilization of continuous-wave (cw) lasers to the level of $1 \times 10^{-15}$ for integration times of 0.1s to 100s or better [19-23]. The combination of these two technologies provides the potential for a compact, simple and extremely low-noise microwave source based on phase-locking one of the frequency-comb modes to the ultra-stable continuous wave laser. Once phase locking is realized, the optical pulse train is detected by a high speed photodiode to extract a microwave signal at a harmonic of the pulse repetition rate.

Unfortunately, the detection and demodulation of the optical train suffers from various excess noise processes which limits the residual phase noise of the optical-to-microwave frequency division process alone. For frequency combs based on Titanium:sapphire (Ti:S) lasers a residual phase noise of -98 dBc/Hz at 1 Hz Fourier frequency offset from a 10 GHz carrier frequency was demonstrated [24]. Furthermore, by implementing an ultra-low noise phase detection system, the residual phase noise at 10 GHz carrier was improved to be -110 dBc/Hz at 1 Hz offset [25]. Although both results were a breakthrough for optical-to-microwave frequency division, they were still limited by excess phase noise which was attributed to the variation of the laser power incident on the photodiode *via* amplitude-to-phase (AM-to-PM) conversion processes in the photodiode [26-28]. Very recently, two independent ultra stable lasers combined with two Titanium:sapphire (Ti:S) frequency combs have shown an absolute phase noise of about -104dBc/Hz at 1 Hz Fourier frequency offset from a 10 GHz carrier frequency [29].

In contrast to frequency combs based on Ti:S, fiber-based optical frequency combs (FOFCs) demonstrate a reliability and repeatability of phase noise performances which makes them much more suited to applications in which they need to operate continuously and with a specified level of performance. As one example, they have been used as the frequency reference for interrogating primary caesium fountain frequency standards over the long-term [2,3]. We recently made use of two near-identical FOFCs, which were phase-locked to a common ultra-stable laser source at 1.55 μm, to demonstrate a residual phase noise of -111 dBc/Hz at 1 Hz offset from a 11.55 GHz carrier frequency [30]. Although essentially equivalent in performance to the Ti:S results, the long term operation and stable performance of the FOFC means that it is much more widely applicable. Nonetheless, the result of that time was limited, at Fourier frequencies close to the carrier, by the noise measurement system including the photodetection, amplification and phase detection stages respectively.

The main objective of this article is a detailed analysis of our work published in [31], together with some additions and improvements to the techniques which result in improvements to the performance. In particular, we investigate the noise sources that limit the performance of the FOFC. This is only possible because of the construction of a new ultra-low noise carrier-suppression noise measurement system which circumvents the phase detection and amplification noise mentioned above. The carrier-suppression technique allows us to directly identify each of the noise sources that limits the performance of the comb as a frequency divider, and then to enable implementation of noise reductions strategies for each of these sources.

We address the impact of the Relative Intensity Noise (RIN) on the photo detection noise floor and apply mitigation strategies by two different power control techniques. We also demonstrate a novel phase noise reduction method based on feed-forward correction which circumvents the limited correction bandwidth available on the laser comb.

As a result the phase noise is now reduced to -120 dBc/Hz at 1 Hz from the 11.55 GHz carrier. We believe that the residual noise floor at 1Hz Fourier frequency is within a factor of 10 of the detection shot-noise limit, while it sits within a factor of 4 of this limit from $10-10^7$ Hz. The $1/f$ excess noise is believed to be associated with the detector itself and is consistent with previous measurements of similar detectors driven by cw light in references [32, 33].

2. Optical-to-microwave synthesis based on FOFC

Fig. 1 shows the experimental setup for transferring the optical cw reference signal into the microwave frequency domain [4, 30]. The optical reference is provided by a cw fiber laser that has been frequency stabilized on an ultra-stable vibration insensitive optical cavity [23, 35] by the Pound-Drever-Hall (PDH) method. This ultra-stable continuous-wave laser provides 4 mW of optical power at about 1542.14 nm wavelength ($v_{cw}$). Using the technique described below, one mode of an erbium-doped fiber optic frequency comb (FOFC) with 250 MHz pulse repetition rate ($f_{rep}$) is phase-locked to the ultra-stable laser. This FOFC also contains a built-in f-2f interferometer for generating the carrier-envelope offset frequency: $f_0$. Our technique eliminates the effect of $f_0$ which has the effect of ensuring that the repetition rate of the output pulse train is phase-coherent with the optical reference. In the frequency domain, the FOFC spectrum is composed of many evenly spaced lines of frequencies $v_n = n f_{rep} + f_0$ with $n$ being a (large) integer referencing the line of frequency $v_n$. The FOFC characteristics can be adjusted by two separate actuators: a relatively low bandwidth piezo-electric stretcher that changes the cavity length and thereby principally changes $f_{rep}$, and secondly by changing the pump current, which gives higher speed control of $f_{rep}$ through complex laser dynamics.

In the experiment, the frequency comb output optical power is about 30 mW over a spectral range from 1470 nm to 1620 nm (the -20 dB bandwidth). After passing through a polarization controller, the comb is combined with the ultra-stable laser output using a three-port coupler, whose output is coupled to an optical add/drop multiplexer (OADM in Fig.1). The OADM is a low-loss three-port device which spectrally splits the input light into two parts: a first narrow slice of 0.8 nm (100 GHz) centered at 1542.14 nm while the other output contains the rest of the optical spectrum. The output consisting of the ultra-stable laser and a few adjacent comb lines around 1542.14 nm is detected by an InGaAs photodiode (PD) having a bandwidth of 2 GHz. This PD produces an ensemble of beatnote signals between the cw laser and the nearest frequency comb modes.

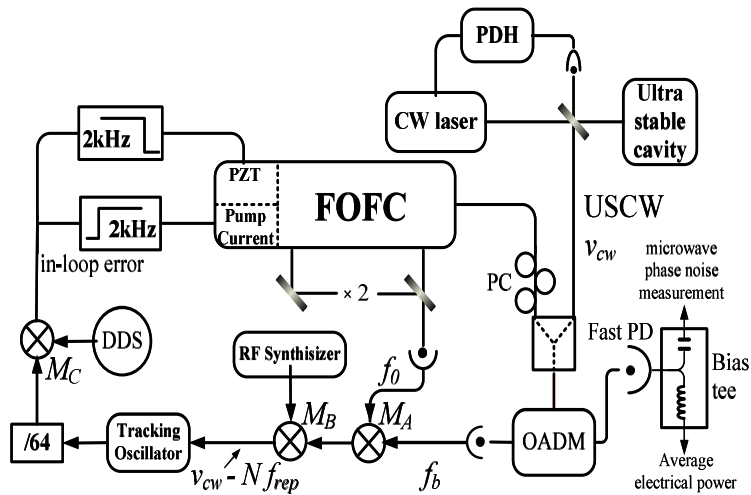

Figure 1 Experimental setup. FOFC: fiber-based optical frequency comb; USCW: ultra-stable continuous-wave laser; PDH: Pound-Drever-Hall; PC: polarization controller; OADM: optical add/drop module; DDS: direct digital synthesizer.

We choose one of these, whose frequency is $f_b = v_{cw} - N f_{rep} - f_0$ (with $N$ a large integer), which we manually adjust to set its frequency close to 428 MHz by acting on the FOFC PZT actuator. The FOFC is also independently adjusted to set its carrier-envelope offset frequency $f_0$ (produced by the built-in f-2f interferometer) near 20 MHz. By mixing (mixer A) these two RF signals together and then remixing (mixer B) the product with another reference at 268 MHz (produced by RF synthesizer B), we obtain a signal at a frequency near 140 MHz which is independent of $f_0$ [33]. A home-made tracking oscillator filter is then used to eliminate AM noise and serve as an agile narrowband filter near 140 MHz. It works by phase-locking a voltage controlled oscillator to the input beat (passed through a logarithmic amplifier beforehand) by use of a double-balanced mixer as a phase detector. The bandwidth of this tracking oscillator phase-lock is about 1 MHz. This clean output signal is then shifted up to 860 MHz by mixing it with a 1 GHz reference before it is digitally frequency divided by a factor 64. The up-shift in frequency before division is necessary for driving the digital divider in its optimal operation window. On the other hand, the frequency division stage is necessary to ensure reliability of the phase lock between the optical comb mode and the cw laser reference. In order for any phase-lock to operate reliably it is a requirement that the integrated residual phase noise within the phase-locked loop bandwidth is small compared with 1 rad rms. In this experiment the actuator bandwidth on the comb has a relatively narrow bandwidth (10 kHz) and it thus requires a reduction in the free-running phase noise which is delivered by the frequency division stage.

After the frequency division, we mix again with a Direct Digital Synthesizer (DDS) reference at 13.4375 MHz to create (after low pass filtering) an error-signal suitable for correction of the FOFC's repetition rate. The error signal passes through a dual-stage loop filter to optimize the noise suppression. In one path the error signal passes through a proportional, double stage integrator, and phase advance analog circuit and is then applied to the FOFC's pump current control. In the second path we integrate this signal again and it then drives the FOFC's PZT. As a result, phase noise at Fourier frequencies above 2 kHz is effectively corrected by the pump control of the FOFC while Fourier frequencies below 2 kHz are taken care of by the PZT. When this phase-lock loop is on, $f_{rep}$ is coherently linked to $\nu_{cw}$ and (within the bandwidth of the phase-lock loop) ideally copies its spectral purity.

To generate $f_{rep}$ and its harmonics, a fiber-coupled 20 GHz bandwidth PIN InGaAs photodiode (DSC40, Discovery Inc.) is connected to the OADM port containing the major part of energy of the FOFC output (only comb lines near 1542 nm have been removed). The typical optical power impinging on the photodetectors is about 12mW. The frequency of interest through out this paper is 11.55 GHz, which is the 46$^{th}$ harmonic of fundamental $f_{rep}$. It's typical power level is -30 dBm. The near DC and AC outputs of the fast PD are separated by a bias-Tee as shown in Fig. 1. The RF and microwave part up to 20 GHz is processed by the phase noise measurement system while the average power (ie the near DC signal) is measured from the low frequency output.

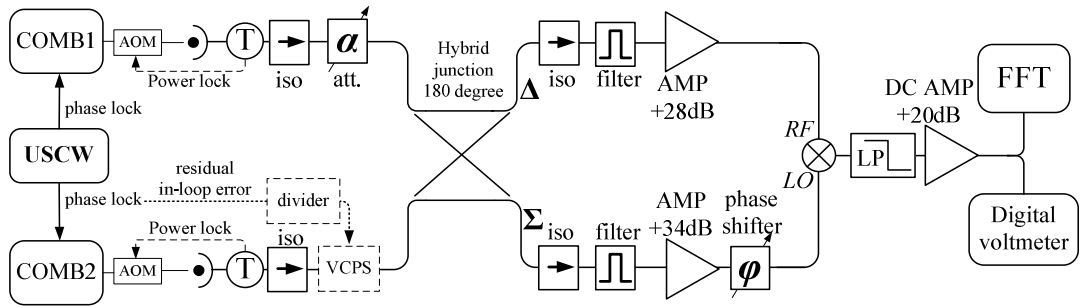

Figure 2. Carrier-suppression noise measurement system. T: bias- tee; Iso: isolator; α: attenuator; filter: band-pass filter at 11.55GHz; AMP: low-noise amplifier; ϕ: phase shifter; LP: 50MHz low-pass filter; DC AMP: low noise DC amplifier; FFT: Fast Fourier Transform analyzer; VCPS (in dash line): voltage control phase shifter; Divider (in dash line): voltage divider.

To characterize the residual phase noise of optical-to-microwave frequency division process, a second fiber-based frequency comb with quasi-identical structure and equivalent locking scheme is phase-locked to the same ultra-stable laser. The one relevant difference in the two frequency combs used in this experiment arises from differences in the pump current drivers which sets an upper limit to the bandwidth of the FOFC phase locked loop. One comb, termed Comb 1 has a bandwidth of about 150 kHz while the other, Comb 2, is only 20 kHz.

*3. Carrier-suppression scheme for low phase noise measurements*

In ref. [30], we have demonstrated a 11.55 GHz microwave signal generated by a FOFC with a residual phase noise of -111 dBc/Hz at 1 Hz offset frequency (for two systems). However, this measurement was based on a traditional phase comparison scheme using a double balanced mixer (DBM) and would be limited at low Fourier frequencies by flicker noise originating from the mixer itself. Furthermore, to reach adequate input powers on this mixer it was necessary to employ a number of amplification stages as we start from only -30 dBm from the PD output at the frequency of interest. These amplifiers also contribute to the flicker noise of the measurement and give a white thermal noise induced phase noise floor.

To investigate directly the residual phase noise of the optical-to-microwave division process in fiber frequency combs, we implemented a carrier suppression noise measurement system (CSNMS) to circumvent these noise sources in the phase comparison scheme [37,38]. As shown in Fig. 2, the CSNMS scheme consists of two FOFCs (described in Section 2) which have been phase-locked to a common ultra-stable laser. Each comb uses its own fast PDs for microwave signal generation. Their outputs are then combined in a 4-port 180 degree hybrid junction (bandwidth 6 GHz-12 GHz), producing the sum and the difference of the input signals. A microwave variable attenuator before one of the hybrid junction inputs (shown in Fig. 2) allows amplitude equalization of the microwave signals. Optimal carrier suppression is obtained by adjusting the time of arrival of the light pulses on the photodetectors. Changing the RF synthesizer frequency (LO input of mixer B, Fig. 1) in one of the comb phase-locked loop allows tuning of the time delay between the microwave domain electric signals generated by the two fast PDs.

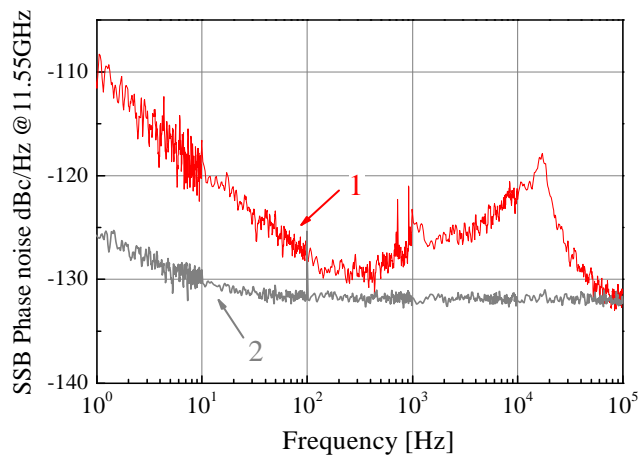

Fig. 3. Relative Residual phase noise of the optical-to-microwave division process (Curve 1). Curve 2: readout noise floor of the carrier suppression noise measurement system

Once time of arrival and amplitude matching is achieved, the signal is suppressed on the difference port of the hybrid junction, thereby producing a *simultaneous* quasi-extinction of all the harmonics of $f_{rep}$ (at least within the 6 GHz-12 GHz bandwidth of the hybrid junction). By optimizing phase and amplitude matching, we can obtain more than 70 dB of carrier suppression at the frequency of interest, resulting in less than -100 dBm signal level on the difference port at 11.55 GHz before amplification. The carrier suppression shows excellent stability without the need for any automatic or manual control over time scales of many hours.

The difference port output is then filtered around 11.55 GHz by three low-insertion loss narrow bandwidth microwave cavity filters (about 1.5 dB insertion loss in each) before being amplified. Flicker phase-noise is well known to arise in non-linear microwave devices due to high driving powers. Therefore, the very low-level of the carrier-suppressed signal means that the flicker noise added by the following low noise amplifier (28 dB gain) and double-balanced mixer is significantly reduced [37,38]. To demodulate the weak signal from the difference port with strong carrier suppression, the signal from the sum port is also filtered near 11.55 GHz and then amplified to about +7 dBm by two amplifiers with an other cavity filter (with total net gain of 34 dB). It is then used to drive the LO port of the mixer which takes the carrier suppressed signal on its RF port. A microwave phase shifter shown in the LO arm of the mixer is used to set the mixer to be sensitive to either phase or amplitude fluctuations. The mixer output is low-pass filtered, amplified by a low-noise DC amplifier and then sent to a Fast Fourier Transform (FFT) analyzer for the phase noise measurement.

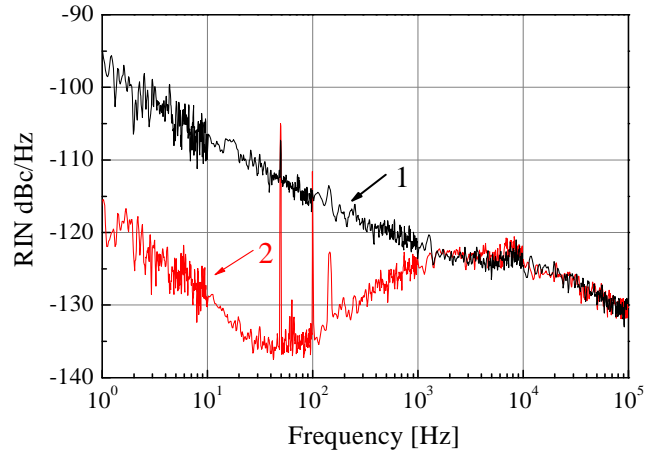

Fig. 4. Curve 1: Relative intensity noise (RIN) of the femtosecond comb. Curve 2: RIN with acousto-optic-modulator-based power lock

The calibration of the sensitivity in the carrier suppression method is a crucial issue that we have addressed by implementing three independent methods. The first method is based on a Voltage-Controlled Phase Shifter (VCPS) with known sensitivity and inserted after one fast PD and biased in the linear regime. The CSNMS is tuned close to the carrier-suppression point and the VCPS is modulated by a random white noise from the signal source of a 2-channel FFT. The known sensitivity of the VCPS is combined with the measured FFT transfer function to generate the frequency-dependent transfer function of the CSNMS. The second calibration method is to replace the mixer LO input by a signal coming from a slightly frequency detuned microwave synthesizer signal, with identical amplitude. We monitor the modulated beat note signal at the mixer output and we calculate the phase sensitivity in the same way one would do in a conventional system.

The third and most convenient way to calibrate the CSNMS is to modulate the optical pulse time of arrival by applying a frequency modulation to the RF synthesizer (268 MHz) shown driving mixer B in Fig. 1. The microwave phase modulation corresponding to this optical phase change can be calculated as $\phi_{\mu W}=2\pi \times \Delta\phi_{opt} \times \nu_{\mu W}/\nu_{opt}$, where $\Delta\varphi_{opt}$ is the optical phase change due to modulation on RF synthesizer, $\nu_{opt}$ is the optical frequency at 194 THz and $\nu_{\mu W}$ is the microwave frequency at 11.55 GHz. Consequently, the CSNMS voltage to phase conversion factor $k_d$ is equal to the peak-to-peak voltage value of the mixer output divided by the microwave phase change calculated above. The voltage-to-phase conversion factors obtained using the three calibration methods differ by less than 0.2 dB.

After setting and calibrating the CSNMS at phase sensitivity point, we evaluate the readout noise floor, measured by replacing the RF input of the mixer by a 50 Ω termination to mimic the condition of perfectly destructive interference. The readout noise floor reported in Fig. 3, starts from -125 dBc/Hz at 1 Hz Fourier frequency offset from carrier frequency and decreases down to a white phase noise floor at -132 dBc/Hz after 40 Hz offset frequency.

As a consistency check, we have measured the residual phase noise of the microwave signal at 11.55 GHz using the CSNMS scheme, which is in agreement within 2-3 dB of our previous result [30] based on a classical phase noise measurement scheme (Fig. 3).

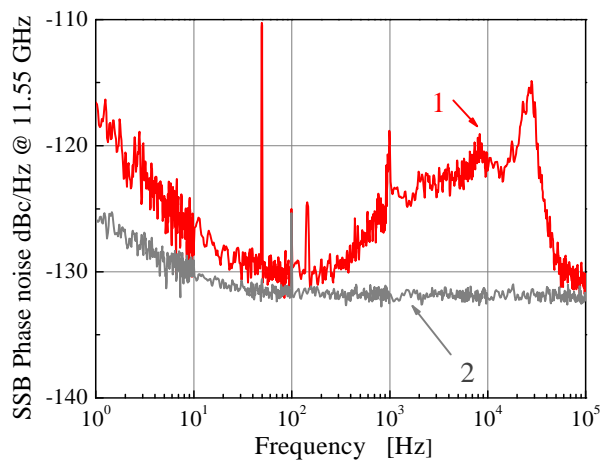

Fig. 5. Curve 1: Residual phase noise of the optical-to-microwave division process with the acousto-optic-modulator-based power lock for near-carrier excess phase noise reduction. Curve 2: readout noise floor of the carrier suppression noise measurement system.

*4. Relative intensity noise measurement and AOM-based power lock*

The excess 1/*f* phase noise seen on the residual phase noise (Curve 1, Figure 3) in the 1 Hz to 200 Hz range is likely to be due to amplitude-to-phase conversion (AM-to-PM) [26-28] during the photodetection process. By this mechanism, the optical power fluctuations of the laser pulse received on the PD change the shape of the detected microwave output pulse, leading to an effective temporal shift, and hence to a microwave phase shift at a given frequency of interest. We can define an AM-to-PM conversion factor to relate the relative intensity noise (RIN) of laser to the phase noise. We express it in the units of [rad/(W/W)].

We measured the RIN of both combs (only one is shown in curve 1, Fig 4 but they were nearly identical) using the low frequency output of the bias-T connected to the PD. The values reported in the literature for the AM-to-PM factor are in the range 0.1-0.5 rad for a saturated PIN photodiode of similar type to that used here [39].

The measured RIN PSD multiplied by such coefficient (~ $(0.3 rad)^2$, i.e.~-10 dB) corresponds closely to the level of measured $1/f$ residual phase noise (1 Hz-100 Hz, in curve of figure 3, Curve 1) and suggests that we would benefit from actively stabilizing the optical power impinging on the fast PD. The optical power incident on the fast PD is controlled by a fiber-coupled acousto-optic modulator (AOM). By acting on this AOM, the average output voltage of the bias-tee is actively stabilized at a constant value. In practice, the error signal is integrated and applied to the amplitude-modulation port of the RF synthesizer that is driving the AOM.

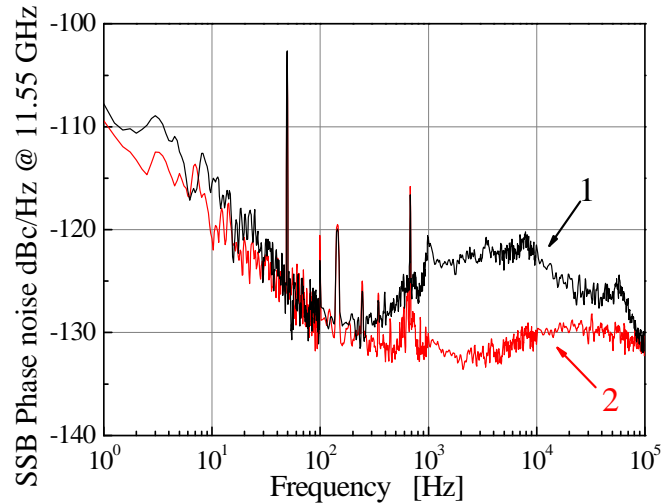

Fig. 6. Curve 1: Single-side-band (SSB) phase noise power spectral density measured without phase noise correction. Curve 2: voltage-controlled phase shifter into low bandwidth fiber-based optical frequency comb for residual feed-forward phase noise correction.

To measure independently the efficiency of the power stabilization, we insert a fibre-splitter and use a second fast PD on the second port of the splitter. The power spectral density (PSD) of the residual RIN measured with this out-of-loop photodiode is shown in Fig. 4 (curve 2). A noise reduction of about 20 dB at 1 Hz offset frequency is obtained. We limit the bandwidth of the power lock scheme to about 1-2 kHz because, at higher Fourier frequencies, the RIN is not the dominant source of phase noise. We implement an AOM-based power lock on both frequency combs (black dashed lines in Fig. 2) and then measure the residual phase noise of the optical-to-microwave division process (see Fig. 5). Curve 1 (red line) with power-lock shows a noise reduction of 6-7 dB at 1 Hz compared to results presented in Ref. 30. We believe that the low frequency phase noise is no longer dominated by residual AM since the AM has been suppressed by a larger factor than the observed phase noise reduction.

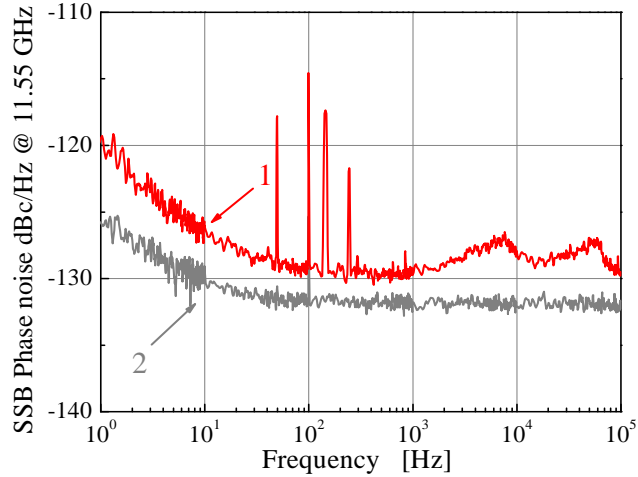

Fig. 7. Single-Side-Band (SSB) phase noise power spectral density of microwave signal at 11.55GHz measured by the Carrier Suppression Noise Measurement System (CSNMS). Curve 1: voltage control phase shifter and pump-current-based power lock scheme applied for phase noise reduction. Curve 2: readout noise floor of the CSNMS.

*5. Voltage-controlled phase shifter for high bandwidth phase-correction*

Despite the success of the power control for reducing excess phase noise associated with power fluctuations, we note that there is still a large phase noise excess between 300 Hz and 80 kHz Fourier frequencies. This noise actually originates in one of the FOFCs (comb 2) whose PLL bandwidth is limited to 20 kHz (i.e. at high Fourier frequencies the intrinsic phase fluctuations of the comb have not been fully suppressed by the phase lock to the ultra-stable laser). To suppress the phase noise in this low control bandwidth regime, we implement a feed-forward noise reduction technique using a VCPS (dashed line in Fig. 2) to apply a phase correction directly to the electronic signals exiting the photodiode. The VCPS operates in its linear range (voltage to phase conversion coefficient is constant) and is inserted after the fast PD of comb 2. The VCPS driving signal is derived from the residual in-loop error (*i.e.* the output of mixer *Mc*). A passive resistive network with an adjustable voltage division factor is connected between the mixer output and the VCPS input. Its dividing ratio is determined by the following relation

$$1/K_{IF} \times R \times Kvcps = (194 \text{ THz}/64)/11.55 \text{ GHz}$$

where $K_{IF}$ is the phase-to-voltage conversion of the mixer *Mc*, *Kvcps* is the VCPS voltage-to-phase factor and *R* is the division ratio of the resistive network. This equation describes the phase ratio between the optical reference and the microwave frequency down-converted by the FOFC, where the factor 64 corresponds to the division ratio of the digital frequency divider in the phase lock loop of Fig. 1.

Figure 6 gives our phase noise measurements with and without VCPS phase correction, respectively. Comparing the two curves in this graph, excess phase noise (300 Hz to 80 kHz) due to the low bandwidth PLL of comb 2 can be reduced by at least 10 dB using the VCPS. This technique is the analog of the "transfer oscillator" scheme developed at the Physikalisch-Technische Bundesanstalt (PTB) [34]. One thing to note is that the VCPS-based phase correction approach exhibits the same limitations of all feed-forward approaches i.e. it needs very low-noise electronic design together with a high linearity and stability of its transfer function. Otherwise it will add noise or drift from the optimum operation point. In addition, due to insertion losses introduced by the VCPS, the noise floor of the CSNMS is slightly increased by about 2 dB. We also observe that the phase noise spectral density shape beyond a few kHz is slightly depending on the environmental and mode locking conditions (Curve 1 Fig. 5 vs Curve 1 figure 6).

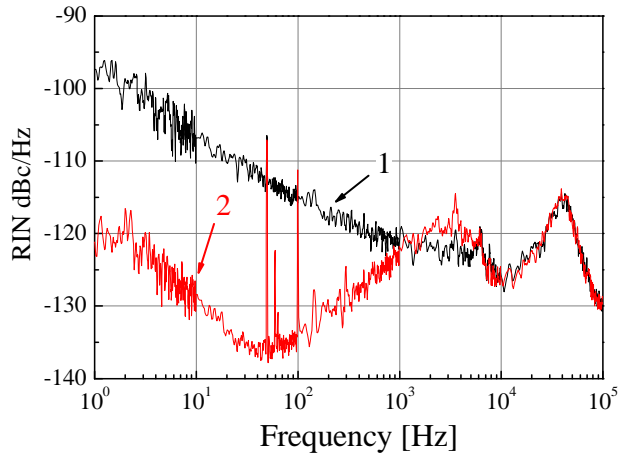

Fig. 8. Relative intensity noise (RIN) measurement based on the out-of-loop fast photodiode. Curve 1: RIN with frequency lock. Curve 2: RIN with frequency lock and pump-current- based power lock.

## 6. Pump-current-based power lock

Though the AOM-based power lock scheme provides good noise reduction in the close to the carrier region, we develop an additional power stabilization scheme by controlling directly the laser-oscillator pump current controller. We choose to undertake this approach for four reasons: (a) it's simpler because we can remove an auxiliary AOM; (b) there is more available power since we remove the insertion loss of the AOM (about 3 dB); (c) for low frequencies (up to 3 kHz) the pump control input was essentially un-used (for these frequencies the repetition rate control in the laser is controlled by the PZT) and only ~1 kHz bandwidth is needed for the intensity control system; (d) there is some concern that even if the power incident on the photodiode is tightly controlled, other components of the fiber circuit may exhibit some amplitude-to-phase conversion. Therefore, by controlling the power in the oscillator itself one might expect additional benefits.

Great care has been paid to avoid coupling between the repetition rate phase-locked loop and the power control system, which in this scheme are applied on the same actuator

(although in different frequency domains). Independent out-of-loop RIN measurements (not shown here) demonstrate that the pump-current-based power lock is restricted to Fourier frequencies below 1 kHz and provides more than 20 dB of noise reduction at 1 Hz.

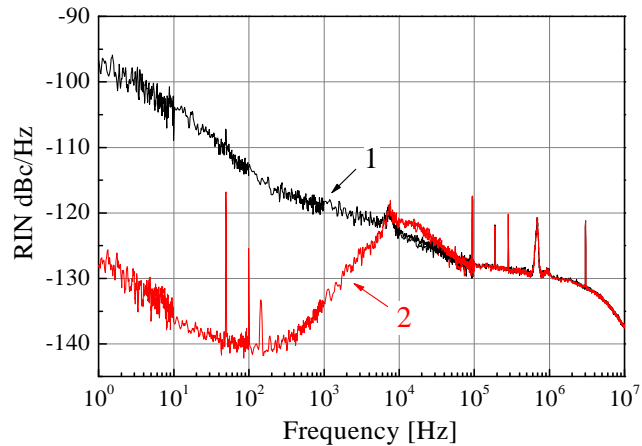

Fig. 9. Curve 1: Relative Intensity Noise (RIN) without power stabilization. Curve 2: RIN with pump-current-based power lock.

We implemented two pump-current-based power lock schemes and measured the residual phase noise while also using VCPS phase correction for the low bandwidth FOFC. As shown in Fig. 7 curve 1, we get slightly better results than that of the AOM-based power lock of figure 5 leading to -120 dBc/Hz at 1Hz. The bump in the phase noise at about 100 kHz is attributed to a dual-step process: the generation of RIN during the frequency correction by the pump current, and then the conversion of this spurious AM into PM during the photodetection. The frequency control using pump current, although appealing, can degrade the high frequency phase noise spectrum.

In a last experiment, to avoid this FM-to-AM-to-PM noise conversion, we rely solely upon the PZT for phase locking (which implies a control bandwidth below 20 kHz) and then made use of only the VCPS feed-forward correction scheme above that frequency. We therefore equip both systems with VCPSs. This approach has the secondary advantage that the pump current is now fully available for the power control with much higher bandwidth than before. In this last configuration we investigate the laser RIN and phase noise up to 10 MHz Fourier frequencies.

The RIN measurement based on the out-of-loop fast PD measurement scheme is shown in Fig. 9. No bump is now observable in the RIN PSD and the free-running RIN is suppressed for frequencies below ~10 kHz (curve 2 in Fig. 9).

Fig. 10 shows the residual phase noise measurement under these conditions. Below 100 kHz the level is compatible with our previous results. The near carrier-frequency excess phase noise is reduced to -120 dBc/Hz at 1 Hz by the pump-current-based power lock. The large excess phase noise from ~200Hz to 100 kHz is greatly suppressed by the two VCPS. The measurement however shows an unexplained excess phase noise which rolls-off after 500 kHz and reaches the readout noise floor around 3 MHz. This peculiarity of the noise spectrum seems to be related with the mode-lock state of the femtosecond laser comb as its level varies when the comb is mode-locked in a different state.

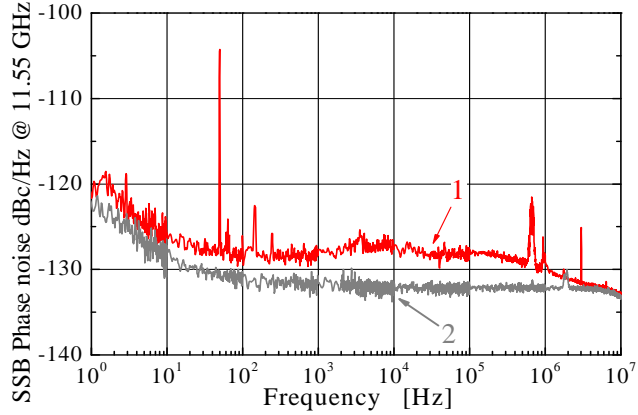

Fig. 10. Curve 1: pump-current-based power lock for near-carrier excess phase noise reduction and two voltage control phase shifters for phase correction. Curve 2, readout noise floor of the carrier suppression noise measurement system.

*Conclusions*

We have demonstrated an ultralow residual phase noise optical-to-microwave frequency division process using fiber-based optical frequency combs and advanced noise reduction techniques. A microwave signal at 11.55 GHz, corresponding to the $46^{th}$ harmonic of the frequency comb repetition rate, is extracted from two frequency combs referenced to a common ultra-stable laser. We implemented a carrier-suppression noise measurement system to reduce greatly the noise floor of the microwave phase comparison. This ultra-low noise system allowed us to directly identify several sources of noise which prevented optimal performance of the optical frequency division process. We then implemented two optical power stabilization schemes in order to suppress the effect of the AM-to-PM noise conversion. This stabilization provided effective noise suppression at low Fourier frequencies, namely -117 dBc/Hz at 1 Hz in a AOM-based power lock scheme and -120 dBc/Hz at 1Hz from pump-current-based power lock scheme. The residual noise with its flicker-character is thought to be associated with the intrinsic characteristics of fast photodiodes. At high Fourier frequencies the residual phase noise associated with the limited bandwidth of the frequency actuator in one of the frequency combs is reduced by 10 dB using a feed-forward technique with a voltage-controlled phase shifter.

Further improvements on ultralow phase noise microwave signals generation would require the use of photodiodes with a lower flicker noise to reduce excess phase noise in the near-carrier Fourier frequency region. Alternatively the use of a photodiode that can tolerate high power could increase the signal-to-noise ratio which may improve the white phase noise floor. As a third alternative it would be possible to enhance the signal-to-noise ratio of the output microwave frequency by increasing the frequency comb repetition rate either directly or by using a Fabry-Perot cavity [40, 41] to filter the comb prior to being directed onto the photodiode. The implementation of an intra-cavity electro-optical modulator for repetition rate control with a bandwidth of a few 100 kHz

[42,43], could greatly improve the robustness of the microwave generation, and may remove the need on the VCPS feed-forward technique for efficient noise rejection. Moreover this will make the pump power control fully available for RIN effect mitigation. The preference between AOM based and pump power RIN reduction is not completely determined. An AOMs have potentially faster response, but the pump power control allows in the other hand the stabilization of the laser intracavity power for better long term behavior [31].